\documentclass[11pt,a4paper]{article}
\usepackage{jheppub}
\usepackage{amsmath,amssymb}
\usepackage{slashed}
\usepackage{graphicx}
\usepackage{subfigure}
\usepackage{url}
\usepackage{pstricks}
\newcommand{\beq}{\begin{equation}}
\newcommand{\eeq}{\end{equation}}

\newcommand{\be}{\begin{eqnarray}}
\newcommand{\ee}{\end{eqnarray}}
\long\def\hidestart#1\hideend{}
\title
{Open Boundary Condition, Wilson Flow and the Scalar Glueball Mass}
\author{Abhishek Chowdhury$^{a}$,}
\author{A. Harindranath$^{a}$ and}
\author{Jyotirmoy Maiti$^{b}$}

\affiliation{$^{a}$Theory Division, Saha Institute of Nuclear Physics \\
 1/AF Bidhan Nagar, Kolkata 700064, India}

\affiliation{$^{b}$Department of Physics, Barasat Government College,\\
10 KNC Road, Barasat, Kolkata 700124, India}

\emailAdd{abhishek.chowdhury@saha.ac.in}
\emailAdd{a.harindranath@saha.ac.in}
\emailAdd{jyotirmoy.maiti@gmail.com}

\date{February 24, 2014}
\abstract {
A major problem with periodic boundary condition on the gauge fields used in current
lattice gauge theory simulations is the trapping of topological charge in a particular sector
as the continuum limit is approached. To overcome this problem open boundary condition in the
temporal direction has been proposed recently. One may ask whether open boundary condition
can reproduce the observables calculated with periodic boundary condition.
In this work we find that the extracted lowest glueball mass using open and periodic
boundary conditions at the same lattice volume and lattice spacing agree for the range of 
lattice scales explored in the range $3~{\rm GeV}\leq \frac{1}{a}\leq 5$ GeV.
The problem of trapping is overcome to a large extent with open boundary and we
are able to extract the glueball mass at even larger lattice scale $\approx $ 5.7 GeV.
To smoothen the gauge fields we have used recently proposed Wilson flow
which, compared to HYP smearing, exhibits better systematics in the extraction of
glueball mass. The extracted glueball mass shows remarkable insensitivity to the lattice spacings 
in the range explored in this work, $3~{\rm GeV}\leq \frac{1}{a}\leq 5.7~{\rm
GeV}$.}

\begin{document}

\maketitle

%%%%%%%%%%%%%%%%%%%%%%%%%%%%%%%%%%%%%%%%%%%%%%%%
\section{Motivation}
Even though lattice QCD continues to make remarkable progress in confronting 
experimental data, certain problems have persisted. For example, the spanning
of the gauge configurations over different topological sectors become progressively
difficult as the continuum limit is approached. This is partly intimately related 
to the use of periodic boundary condition on the gauge field in the temporal 
direction of the lattice. As a consequence, in the continuum limit, different
topological sectors are disconnected from each other. Thus at smaller and smaller
lattice spacings the generated gauge configurations tend to get trapped in
a particular
topological sector for a very long computer simulation time thus resulting in very
large autocorrelations. This may sometime even invalidate the results of the simulation.
Open boundary condition on the gauge field in the temporal direction has been 
recently proposed to overcome this problem \cite{open0,open1,open2}.
Usage of open boundary conditions has been found to be advantageous in a
weak coupling study of SU(2) lattice gauge theory \cite{grady}.

The spanning of different topological sectors can be studied through topological 
susceptibility ($\chi$) which is related to the $\eta^{\prime}$ mass by the Witten-Veneziano 
formula \cite{witten,veneziano,seiler} in pure Yang-Mills lattice theory.
For example, some high precision calculations of $\chi$ on periodic lattices
are provided in Refs. \cite{del,durr,lspb}.
In Ref. \cite{opentopo}, we have addressed the question whether open boundary condition
in the temporal direction can yield the expected value of $\chi$. We have shown that
with the open boundary it is possible to get the expected value of $\chi$ and the
result agrees with our own numerical simulation employing periodic boundary condition.

We continue our exploration of open boundary condition in this work, in the context of
extraction of lowest glueball mass from the temporal decay of correlators.
Extraction of glueball masses compared to hadron masses is much more difficult due to
the presence of large vacuum fluctuations present in the correlators of gluonic 
observables. Moreover the computation of low lying glueball masses which are much
higher than the masses of hadronic ground states, in principle requires relatively 
small lattice spacings. To overcome these problems, anisotropic lattices together with
improved actions and operators have been employed \cite{mp1,mp2,chen} successfully
to obtain accurate glueball masses. On the other hand, the calculation of glueball masses
with isotropic lattice has a long history (see for example, the reviews, Refs. \cite{teprev,balirev}).
These calculations which employ periodic boundary condition in the temporal direction
have been pushed to lattice scale of $ a^{-1} = 3.73(6)$ GeV
\cite{baliplb,vaccarino}. One would
like to continue these calculations to even higher lattice scale which however eventually 
will face the problem of efficient spanning of the space of gauge configurations.
Such trapping has been already demonstrated \cite{opentopo}. It is interesting
to investigate whether the open boundary condition can reproduce the glueball
masses extracted with periodic boundary condition at reasonably small lattice
spacings achieved so far and whether the former can be extended to even smaller
lattice spacings. Our main objective in this paper is to address these issues.

An important ingredient in the extraction of masses
is the smearing of gauge field which is necessary both to suppress unwanted fluctuations
due to lattice artifacts and to increase the ground state overlap \cite{parisi}.
In the past various techniques have been proposed towards smearing the gauge fields
\cite{ape,hyp,morningstar}. Recently proposed Wilson flow \cite{wf1,wf2,wf3} puts 
the technique of smearing on a solid mathematical footing. The same idea is referred 
to in the mathematical literature by the name of gradient flow \cite{atbott,gtp,mani}.
Another motivation of the present work is the study of the effectiveness of Wilson flow
in the extraction of masses.

%%%%%%%%%%%%%%%%%%%%%%%%%%%%%%%%%%%%%%%%%%%%%%%%%%
            
%%%%%%%%%%%%%%%%%%%%%%%%%%%%%%%%%%%%%%%%%%%%%%%%%%
\section{Simulation details}
%%%%%%%%%%%%%%%%%%%%%%%%%%%%%%%%%%%%%%%%%%%%%%%
\begin{table}
\begin{center}
\begin{tabular}{|c|l|l|l|l|l|l|l|}
\hline \hline
Lattice & Volume & $\beta$ & $N_{\rm cnfg}$ &$N_0$ &$\tau$&$a[{\rm fm}]$ & $t_{0}/a^2$\\
\hline\hline
{$O_1$}&{$24^3\times48$}&{6.21} &{3970} &{12} &{3}& {0.0667(5)} & {6.207(15)}\\
\hline
{$O_2$}&{$32^3\times64$}&{6.42} &{3028} &{20}&{4}&{0.0500(4)} & {11.228(31)}\\
\hline
{$O_3$}&{$48^3\times96$}&{6.59} &{2333} &{26}&{5}&{0.0402(3)} & {17.630(53)}\\
\hline
{$O_4$}&{$64^3\times128$}&{6.71} &{181} &{64}&{10}&{0.0345(4)} & {24.279(227)}\\
\hline
{$P_1$}&{$24^3\times48$}&{6.21} &{3500} &{12}&{3}&{0.0667(5)} & {6.197(15)}\\
\hline
{$P_2$}&{$32^3\times64$}&{6.42} &{1958} &{20}&{4}&{0.0500(4)} & {11.270(38)}\\
\hline
{$P_3$}&{$48^3\times96$}&{6.59} &{295} &{26}&{5}&{0.0402(3)} & {18.048(152)}\\
\hline\hline
\end{tabular}
\caption{Simulation parameters for the HMC algorithm. $N_0$ is the number of integration steps, $\tau$ is
the trajectory length and $t_{0}/a^2$ is the dimensionless reference Wilson flow time.
$O$ and $P$ refer to ensembles with open and periodic boundary condition in
the temporal direction.}
\label{table1}
\end{center}  
\end{table} 
 
Using the {\tt openQCD} program \cite{openqcd}, SU(3) gauge configurations 
are generated with open boundary condition (denoted by $O$) at different lattice 
volumes and gauge couplings. For comparison purposes, we have also generated gauge
configurations (denoted by $P$) for several of the same lattice parameters
by implementing periodic boundary condition in temporal direction in the {\tt openQCD} 
package. In table \ref{table1}, we summarize details of the simulation parameters.
    
To extract the scalar glueball mass, in this initial study we use the 
correlator of $\overline{E}$ which is the average of the action density over spatial volume at a
particular time slice given in Ref. \cite{open1}. Since the action is a sum over the plaquettes,
this is similar to the use of plaquette-plaquette correlators which have been used in the 
literature \cite{berg,billorie}. As in the latter case, there is room for operator 
improvement. One may use simple four link plaquette (unimproved) or one may use the 
clover definition of the field strength in the action (improved).

Correlator is measured over $N_{\rm cnfg}$ number of configurations.
The separation by $32$ is made between two successive measurements.
Thus the total length of simulation time is $N_{\rm cnfg}\times 32$.
Using the results from Refs. \cite{gsw,necco}, we have determined the
lattice spacings which are quoted in table \ref{table1}.
We have employed Wilson flow \cite{wf1,wf2,wf3} to smooth the gauge configurations.
The implicit equation
\beq
\big\{t^2\langle\overline{E}(T/2)\rangle\big\}_{t=t_0} = 0.3
\eeq 
with $t$ and $T$ being respectively the Wilson flow time and
the temporal extent of the lattice, defines a reference flow time $t_0$
which provides a reference scale to extract physical quantities from lattice
calculations. No significant difference has been found in our results using 
the $w_0$ scale proposed in Ref. \cite{wscale} as an alternative to the $t_0$ scale.

%%%%%%%%%%%%%%%%%%%%%%%%%%%%%%%%%%%%%%%%%%%%%%%%%%
\section{Numerical Results}
%%%%%%%%%%%%%%%%%%%%%%%%%%%%%%%%%%%%%%%%%%%%%%%

\begin{figure}[h]
\begin{center}
\includegraphics[width=3.5in,clip]
{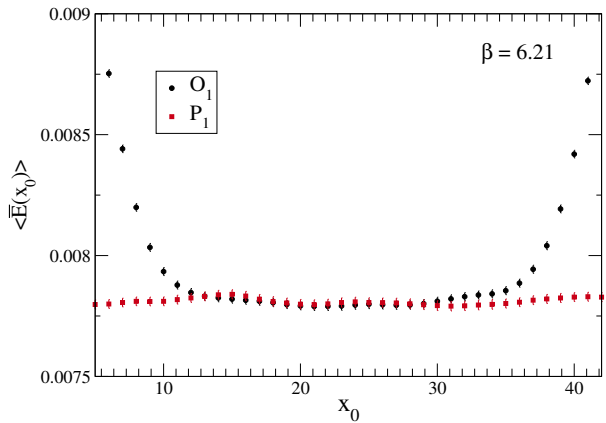}
\caption{Plot of $\langle\overline{E}(x_0)\rangle$ versus $x_0$ at flow time $t=t_0$ at $\beta=
6.21$ and lattice volume $24^3\times 48$ for ensemble $O_1$ (filled circle) and 
 ensemble $P_1$ (filled square).}
\label{edtime}
\end{center}
\end{figure}
Since we extract the scalar glueball mass from the temporal decay  of the 
correlator of $\overline{E}(x_0)$ where $x_0$ denotes the particular
temporal slice,
we first look at the effect of open boundary on the  $\langle\overline{E}(x_0)\rangle$. In figure \ref{edtime}
we plot $\langle\overline{E}(x_0)\rangle$ versus $x_0$ at flow time $t=t_0$ at $\beta=
6.21$ and lattice volume $24^3\times 48$ for ensemble $O_1$. Breaking of translational invariance due to
open boundary condition in the temporal direction is clearly visible in the plot. To calculate the correlator
we need to pick the sink and source points from the region free from boundary artifacts, which can be identified
from such plot. To facilitate the identification better, we also plot $\langle\overline{E}(x_0)\rangle$ for
periodic boundary condition in the temporal direction for the same lattice volume and lattice spacing (ensemble
$P_1$). Preservation of translation invariance is evident in this case. Clearly, for open boundary condition,
source and sink points need to be chosen from the region where  $\langle\overline{E}(x_0)\rangle$ is almost flat. 
We note that for both open and periodic cases the central region  $\langle\overline{E}(x_0)\rangle$ is not 
perfectly flat but exhibits an oscillatory behaviour on a fine scale.

\begin{figure}[h]
\begin{center}
\includegraphics[width=2.8in,clip]
{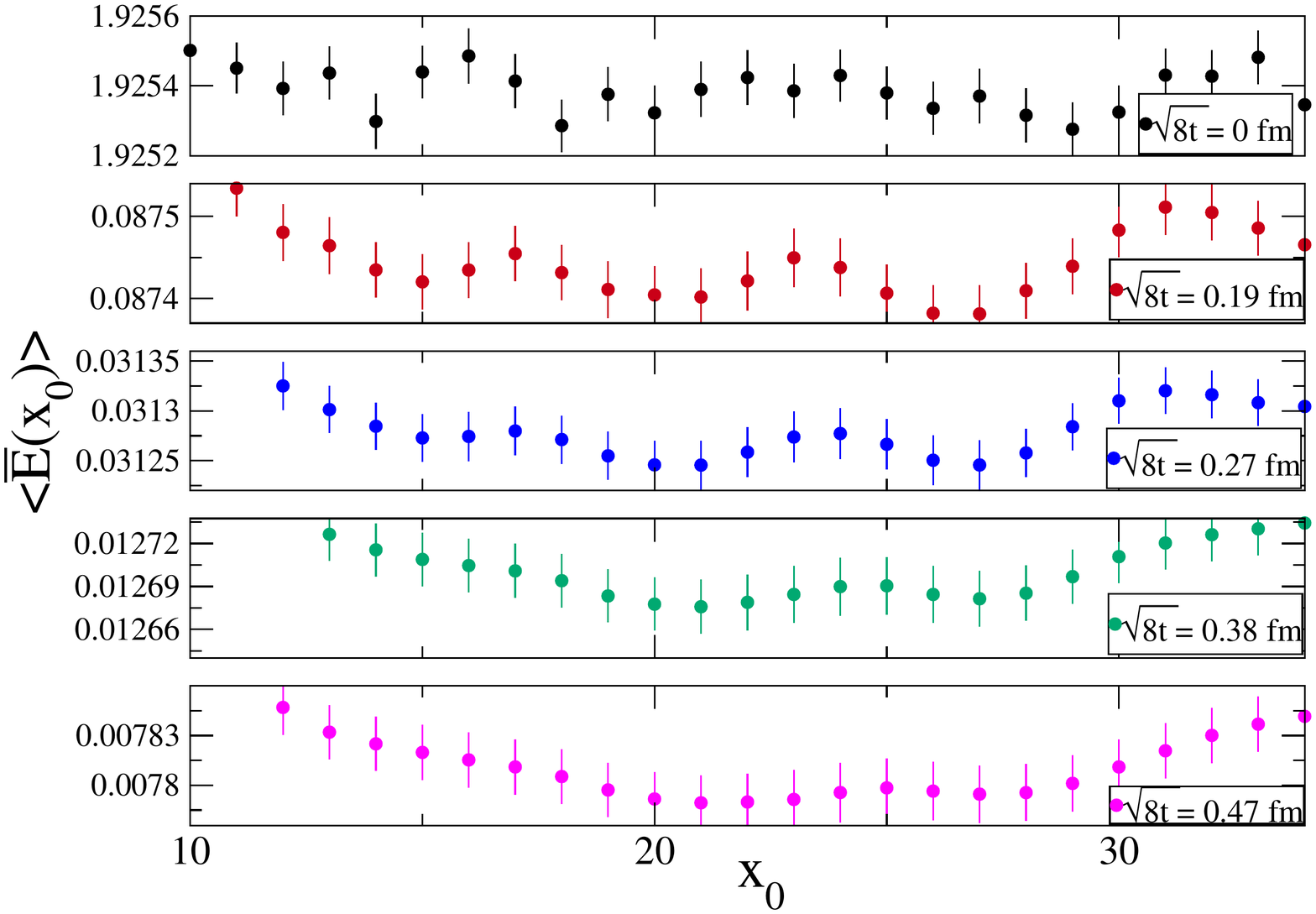}
\includegraphics[width=2.8in,clip]
{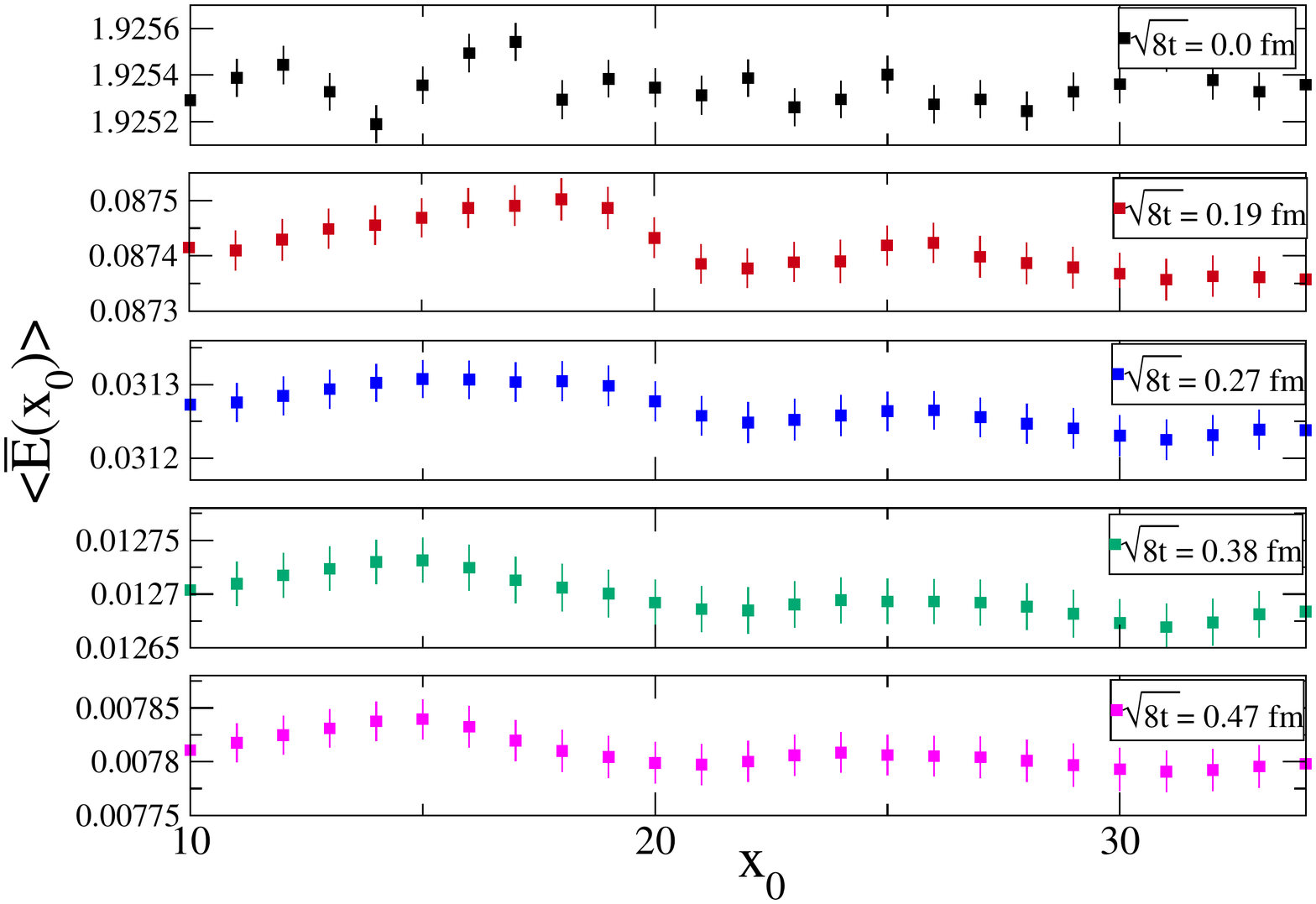}
\caption{Plot of $\langle\overline{E}(x_0)\rangle$ versus $x_0$ at various flow times $t$ at $\beta=
6.21$ and lattice volume $24^3\times 48$ for ensemble $O_1$ (left) and for ensemble $P_1$ (right).}
\label{edfinescale}
\end{center}
\end{figure}

To understand the oscillatory behaviour, in figure \ref{edfinescale} we plot $\langle\overline{E}(x_0)\rangle$ 
versus $x_0$ at various flow times $t$ at $\beta=6.21$ and lattice volume $24^3\times 48$ for ensemble $O_1$ (left) 
and for ensemble $P_1$ (right). At small Wilson flow time, the fluctuations of $\langle\overline{E}(x_0)\rangle$ are
very large as seen from the top panel of the plots. To reduce the fluctuation we have to increase Wilson flow time.
The comparison of different panels clearly demonstrates the reduction of fluctuations with increasing flow time
(note that the scale on y axis becomes finer and finer as flow time increases). However, with increasing flow time 
the data become more correlated and longer wavelengths appear \cite{open2}. The plots show that this smoothening
behaviour is the same for both the open and periodic boundary conditions.

\begin{figure}[h]
\begin{center}
\includegraphics[width=4.5in,clip]
{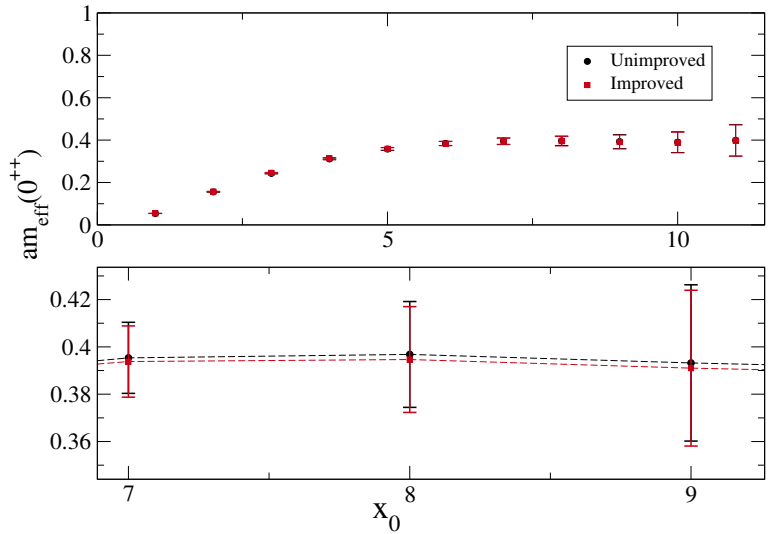}
\caption{Plot of glueball effective mass $am_{eff}~(0^{++})$ versus the
temporal difference $x_0$ at Wilson flow times $\sqrt{8t}=0.28~{\rm fm}$, $\beta=
6.42$ and lattice volume $32^3\times 64$ for ensemble $P_2$ for improved and unimproved choices of operators. The lower panel
shows the detail of the plateau region of the upper panel.}
\label{compWY}
\end{center}
\end{figure}
Next we discuss the extraction of glueball mass. As already discussed in section 2, one may use
the unimproved (naive plaquette) or improved (clover) version of the operator $\overline{E}(x_0)$. 
In general we expect improved operator to be preferable over unimproved one. However, for the extraction of
masses Wilson flow is essential and this may diminish the difference between the results using them.
In this work we have used Wilson flow in all the four directions as originally conceived. Due to the
smearing in the temporal direction we should expect to get glueball mass for separation between source
and sink which are larger than twice the smearing radius ($\approx 2\times\sqrt{8t}$). However a successful
extraction of glueball mass in this case requires reasonably small statistical error at such large
temporal separation.  
In figure \ref{compWY} we plot glueball effective mass $am_{eff}~(0^{++})$ versus
the temporal difference $x_0$ ($x_0 = x_0^{{\rm source}} - x_0^{{\rm sink}}$) at 
Wilson flow time
$\sqrt{8t}=0.28~{\rm fm}$, $\beta=6.42$ and lattice volume $32^3\times 64$ for ensemble $P_2$ for 
improved and unimproved choices of operators (from here onwards, we denote the
temporal difference by $x_0$). As expected the plateau appears for relatively larger 
temporal separation and presumably thanks to Wilson flow the statistical error is reasonably small.
We have verified that the results are very similar at all other
Wilson flow times under consideration. Even though we find that there is no noticeable difference
between them, we employed the improved operator for the rest of the
calculations in this paper.

\begin{figure}[h]
\begin{center}
\includegraphics[width=4.5in,clip]
{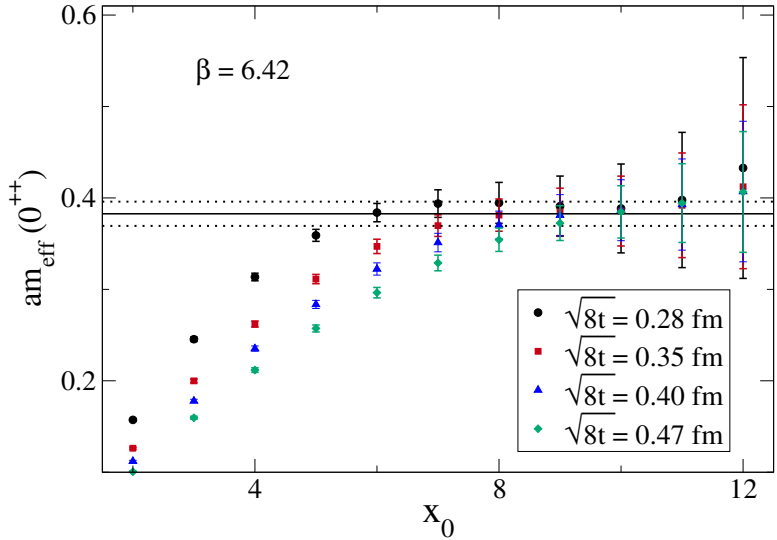}
\caption{Plot of lowest glueball effective mass $am_{eff}~(0^{++})$ versus $x_0$ at
four different Wilson flow times $t$, $\beta=
6.42$ and lattice volume $32^3\times 64$ for ensemble $P_2$. Also shown is
the fit to the plateau region of the data for $\sqrt{8t}= 0.35$ fm.}
\label{smlevel}
\end{center}
\end{figure}
We extract the effective mass for the glueball ($0^{++}$) state from the temporal decay of the correlator 
$\langle\overline{E}(x_0^{{\rm sink}})\overline{E}(x_0^{{\rm source}})\rangle $ where $x_0^{{\rm sink}}$ and 
 $x_0^{{\rm source}}$ are the sink and source points in the temporal direction. To improve the statistics 
we have averaged over the source points when we employ periodic boundary condition on the temporal direction.
Further to reduce fluctuations we have performed the Wilson flow up to flow time $t=t_0$.
In figure \ref{smlevel} we plot the lowest glueball effective mass $am_{eff}~(0^{++})$ versus $x_0$ 
at four Wilson flow times $t$, $\beta=6.42$ and lattice volume $32^3\times 64$ for ensemble $P_2$.
We find that the effective mass is sensitive to Wilson flow time for initial temporal differences $x_0$
 but becomes independent of different Wilson flow times in
the plateau region within statistical error. Note that as expected, the plateau region moves to the
right as Wilson flow time increases. Also shown in the figure is   
the fit to the plateau region of the data for $\sqrt{8t}= 0.35$ fm. The fit
nevertheless passes through the plateau regions of data sets corresponding
to other Wilson flow times.

\begin{figure}[h]
\begin{center}   
\includegraphics[width=4.5in,clip]
{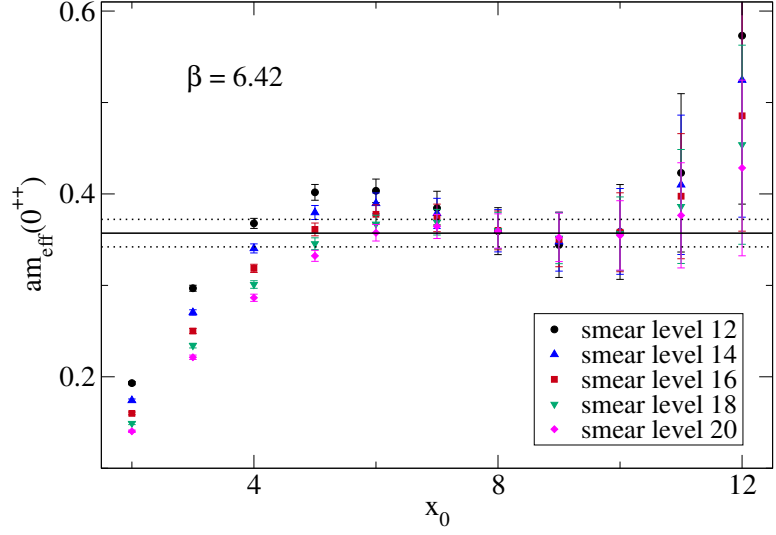} 
\caption{Plot of lowest glueball effective mass $am_{eff}~(0^{++})$ versus
$x_0$ at five HYP smearing levels at $\beta = 
6.42$ and lattice volume $32^3\times 64$ for ensemble $P_2$. Also shown is
the fit to the plateau region of the data for smear level 18.}
\label{HYP}
\end{center}   
\end{figure}
For comparison with traditional methods to smoothen the gauge field
configurations, in figure \ref{HYP} we plot the lowest glueball effective mass
$am_{eff}~(0^{++})$ versus $x_0$
at five smearing levels for four dimensional HYP smearing \cite{hyp} at $\beta=6.42$ 
and lattice volume $32^3\times 64$ for ensemble $P_2$.  
We find that the effective mass for different smear levels
converge in a very narrow window where we can identify the plateau
region and extract the mass. This behaviour is to be contrasted
with that in the case of Wilson flow discussed in the previous
paragraph. Also shown in the figure is   
the fit to the plateau region of the data for smear level 18.
In physical units the fitted mass is found to be 1409 (59) MeV which has
a marginal overlap with the same [1510 (52) Mev] obtained with Wilson flow.
We have observed from our studies with all the $\beta$ values that the results
obtained with HYP smearing are systematically lower than those obtained 
with Wilson flow. We note that the latter value
is closer to the range of glueball mass quoted by other collaborations.
The works presented in the rest of paper employ Wilson flow to smooth the
gauge fields.  

\begin{figure}[h]
\begin{center}
\includegraphics[width=4.5in,clip]
{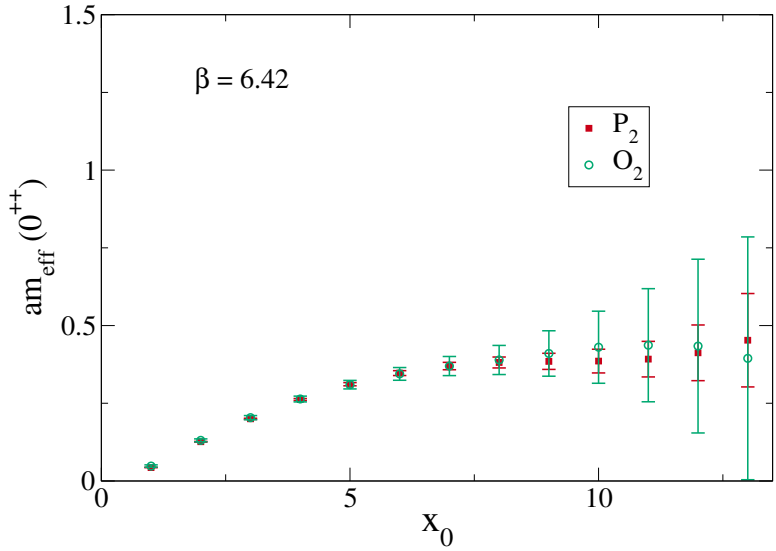}
\caption{Comparison of lowest glueball mass $am_{eff}~(0^{++})$ versus $x_0$ at Wilson flow time 
($\sqrt{8t}=0.35~{\rm fm}$), $\beta=6.42$ and lattice volume $32^3\times 64$ for ensembles $O_2$ and $P_2$.}
\label{compopenpbc}
\end{center}
\end{figure}
With open boundary condition the translational invariance in the temporal direction is broken and 
hence we can not average over all the source points to improve statistical accuracy as we have done
in the case of periodic boundary condition. Nevertheless, we can average over few source points 
chosen far away from the boundary. In figure \ref{compopenpbc} we plot the lowest glueball effective
mass $am_{eff}~(0^{++})$ versus $x_0$ at Wilson flow time ($\sqrt{8t}=0.35~{\rm fm}$), $\beta=6.42$ and lattice 
volume $32^3\times 64$ for both open and periodic boundary conditions (ensembles $O_2$ and $P_2$).
We find that effective mass agree for the two choices of the boundary conditions but as expected
statistical error is larger for open boundary data. 

\begin{table} 
\begin{center}
\begin{tabular}{|c|l|l|}
\hline \hline
Lattice & fit range &$am~(0^{++})$\\
\hline\hline
{$O_1$}&{7-9}&{0.569(69)}\\
\hline
{$P_1$}&{7-9}&{0.520(21)}\\
\hline
{$O_2$}&{9-12}&{0.419(57)}\\
\hline
{$P_2$}&{8-11}&{0.383(13)}\\
\hline
{$O_3$}&{10-12}&{0.327(39)}\\
\hline
{$P_3$}&{10-12}&{0.313(28)}\\
\hline
{$O_4$}&{7-10}&{0.274(48)}\\
\hline\hline 
\end{tabular}
\caption{Lattice glueball $0^{++}$ mass.}
\label{table2}
\end{center}
\end{table}
In table \ref{table2} we have shown the fit range used to extract and the extracted 
lattice glueball mass for the ensembles studied in this paper. A constant is
fitted to extract the mass.

\begin{figure}[h]
\begin{center}
\includegraphics[width=4.5in,clip]
{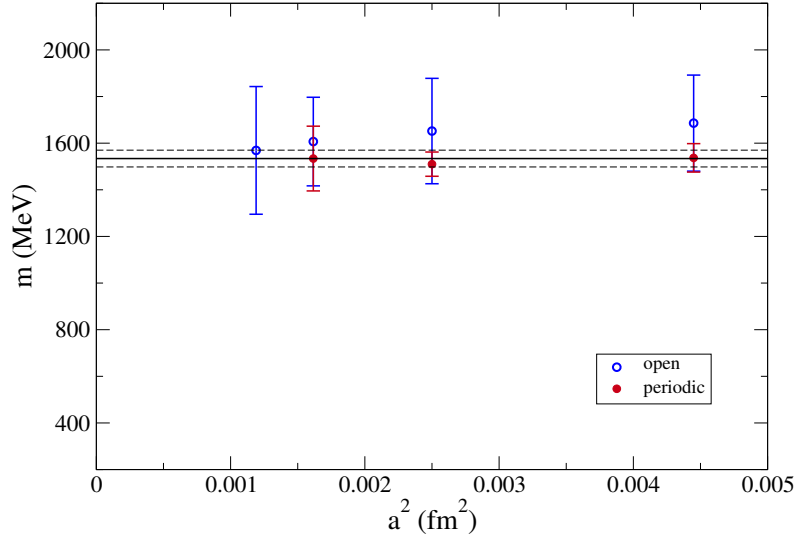}
\caption{Plot of the lowest glueball mass $m~(0^{++})$ in MeV versus $a^2$ for both open and periodic
boundary condition for different lattice spacings and lattice volumes. Also shown is the fit to the combined data.}
\label{gluemass}
\end{center}
\end{figure}
To extract the continuum value of $0^{++}$ glueball mass, in figure \ref{gluemass} we plot $m~(0^{++})$ in MeV 
versus $a^2$ for both open and periodic boundary condition for different lattice spacings and lattice volumes.
%Due to the smoothening effect of Wilson flow, 
For the range of reasonably small lattice spacings explored in this work,   
remarkably, the data does not show any deviation from scaling 
within the statistical error. Hence we fit a constant to the combined data as shown in the figure and extract
the continuum value of $0^{++}$ mass, 1534(36) MeV. We note that this value compares favorably with the 
range of glueball mass quoted in the literature. 

%%%%%%%%%%%%%%%%%%%%%%%%%%%%%%%%%%%%%%%%%%%%%%
\section{Conclusions}
%%%%%%%%%%%%%%%%%%%%%%%%%%%%%%%%%%%%%%%%%%%%%%%%%%%%%%%%%%%%%%%%%%%%%
 In lattice Yang-Mills theory, we have shown that the open boundary condition 
on the gauge fields in the temporal direction of the lattice can reproduce 
the lowest scalar glueball mass extracted with periodic boundary condition 
at reasonably large lattice scales investigated in the range 
$3~{\rm GeV}\leq \frac{1}{a}\leq 5$ GeV.
With open boundary condition we are able to overcome, to a large extent, the problem of trapping
and performed simulation and extract the glueball mass at even larger lattice 
scale $\approx $ 5.7 GeV. Compared to HYP smearing, recently proposed Wilson
flow exhibits better systematics as far as the extraction of glueball mass
is concerned. The extracted glueball mass shows remarkable insensitivity 
to the lattice spacings in the range explored 
in this work $3~{\rm GeV}\leq \frac{1}{a}\leq 5.7~{\rm GeV}$.

Conventionally, due to various theoretical reasons, in the calculation 
of masses from correlators, smearing of gauge field is carried out only in 
spatial directions. In this work, however, Wilson flow is carried out in all 
the four directions and our results show that one can indeed extract mass 
with relatively small statistical error at relatively large temporal 
separations. A critical evaluation of the strengths and weaknesses of the 
four-dimensional versus three-dimensional smoothening of the gauge field in 
the calculation of masses is beyond the scope of the present work.  
%%%%%%%%%%%%%%%%%%%%%%%%%%%%%%%%%%%%%%%%%%%%%%%%%%%%%%%%%%%%%%%%%%%%
\vskip .25in
{\bf Acknowledgements}
\vskip .1in
%%%%%%%%%%%%%%%%%%%%%%%%%%%%%%%%%%%%%%%%%%%%%%%%%%%%%%%%%%%%%%%%%%%%%
Cray XT5 and Cray XE6 systems supported by the 11th-12th Five Year
Plan Projects of the Theory Division, SINP under the Department of 
Atomic Energy, Govt. of India, are used to carry out all the numerical 
calculations reported in this work.
For the prompt maintenance of the systems and the help in data management,
we thank Richard Chang. We also thank Stephan D\"{u}rr and Martin L\"{u}scher for 
helpful comments. This work was in part based on the publicly available lattice gauge theory 
code {\tt openQCD} \cite{openqcd}.

\enlargethispage{\baselineskip}

%%%%%%%%%%%%%%%%%%%%%%%%%%%%%%%%%%%%%%%%%%%%%%%%
%%%%%%%%%%%%%%%%%%%%%%%%%%%%%%%%%%%%%%%%%%%%%%%%%%%%%%%%%%%%%%%%%%%%%%%%   

%%%%%%%%%%%%%%%%%%%%%%%%%%%%%%%%%%%%%%%%%%%%%%%%%%%
\end{document}